\documentstyle[12pt]{article}
\begin{document}
\begin{titlepage}
\begin{flushright}
IC/2001/28\\
hep-th/0105036
\end{flushright}
\vspace{10 mm}

\begin{center}
{\Large The Cardy-Verlinde Formula and \\ 
Asymptotically Flat Charged Black Holes}

\vspace{5mm}

\end{center}

\vspace{5 mm}

\begin{center}
{\large Donam Youm\footnote{E-mail: youmd@ictp.trieste.it}}

\vspace{3mm}

ICTP, Strada Costiera 11, 34014 Trieste, Italy

\end{center}

\vspace{1cm}

\begin{center}
{\large Abstract}
\end{center}

\noindent

We show that the modified Cardy-Verlinde formula without the Casimir 
effect term is satisfied by asymptotically flat charged black holes 
in arbitrary dimensions.  Thermodynamic quantities of the charged black 
holes are shown to satisfy the energy-temperature relation of a 
two-dimensional CFT, which supports the claim in our previous work (Phys. 
Rev. D61, 044013, hep-th/9910244) that thermodynamics of charged black 
holes in higher dimensions can be effectively described by two-dimensional 
theories.  We also check the Cardy formula for the two-dimensional 
black hole compactified from a dilatonic charged black hole in higher 
dimensions.

\vspace{1cm}
\begin{flushleft}
May, 2001
\end{flushleft}
\end{titlepage}
\newpage

Recently, thermodynamics of the holographic duals of various black hole 
solutions have been actively studied, after Verlinde proposed \cite{ver,ver2} 
that the Cardy formula \cite{car} for the two-dimensional conformal field 
theory (CFT) can be generalized to arbitrary spacetime dimensions $D-1$ in 
the form
\begin{equation}
S={{2\pi R}\over{D-2}}\sqrt{E_c(2E-E_c)},
\label{crdvrnlf}
\end{equation}
the so-called Cardy-Verlinde formula \cite{ver,ver2}, where $R$ is the radius 
of the system, $E$ is the total (thermal excitation) energy and $E_c$, 
called the Casimir energy, is the subextensive part of $E$.  (Cf. The quantum 
effects to the Cardy-Verlinde formula were studied in Refs. 
\cite{odi1,odi2}.)  So far, mostly asymptotically AdS black hole solutions 
have been considered \cite{cai,bm,bir,kle1,kle2}.  In Ref. \cite{kle2}, it 
is shown that even the Schwarzschild and the Kerr black hole solutions, which 
are asymptotically flat, satisfy the modification of the Cardy-Verlinde 
formula,
\begin{equation}
S={{2\pi R}\over{D-2}}\sqrt{2E_cE},
\label{zcfwcf}
\end{equation}
to the case of the ground state with zero conformal weight.  In this note, 
we show that this result holds also for various charged black hole solutions 
with asymptotically flat spacetime. 
 
First, we consider the single-charged dilatonic black hole solution with 
an arbitrary dilaton coupling parameter $a$ in arbitrary spacetime 
dimensions $D$.  The action is
\begin{equation}
S_D={1\over{2\kappa^2_D}}\int d^Dx\sqrt{-G}\left[{\cal R}-{4\over{D-2}}
(\partial\phi)^2-{1\over 4}e^{2a\phi}F^2\right],
\label{action}
\end{equation}
where $\kappa_D$ is the $D$-dimensional gravitational constant, $\phi$ is 
the dilaton field and $F$ is the field strength of the $U(1)$ gauge potential 
$A=A_Mdx^M$ ($M=0,1,...,D-1$).  The nonextreme black hole solution to the 
field equations of this action is given by
\begin{eqnarray}
G_{MN}dx^Mdx^N&=&-H^{-{{4(D-3)}\over{(D-2)\Delta}}}fdt^2+H^{4\over{(D-2)
\Delta}}\left[f^{-1}dr^2+r^2d\Omega^2_{D-2}\right],
\cr
e^{\phi}&=&H^{{(D-2)a}\over{2\Delta}},\ \ \ \ \ \ \ \ \ \ \ 
A_t={2\over\sqrt{\Delta}}{{m\sinh\alpha\cosh\alpha}\over{r^{D-3}}}H^{-1},
\label{dilbhsol}
\end{eqnarray}
where
\begin{eqnarray}
H&=&1+{{m\sinh^2\alpha}\over r^{D-3}},\ \ \ \ \ \ \ 
f=1-{m\over r^{D-3}},
\cr
\Delta&=&{{(D-2)a^2}\over 2}+{{2(D-3)}\over{D-2}}.
\label{defs}
\end{eqnarray}
For special values of $\Delta$, this solution can be realized in string 
theories, e.g., the $\Delta=4/n$ case with a positive integer $n$ can be 
obtained by compactifying intersecting $n$ numbers of branes with equal 
charges.  The $a=0$ case is the Reissner-Nordstr\"om black hole.  
The ADM mass $M$, the Bekenstein-Hawking entropy $S$, the Hawking 
temperature $T$, the electric charge $Q$, and the chemical potential 
$\Phi$ of the solution are
\begin{eqnarray}
M&=&{V_{D-2}\over{2\kappa^2_D}}\left[{{4(D-3)}\over\Delta}m\sinh^2
\alpha+(D-2)m\right],
\cr
S&=&{{4\pi V_{D-2}}\over{2\kappa^2_D}}m^{{D-2}\over{D-3}}\cosh^{4\over
\Delta}\alpha,\ \ \ \ \ \ \ 
T={{D-3}\over{4\pi}}m^{-{1\over{D-3}}}\cosh^{-{4\over\Delta}}\alpha,
\cr
Q&=&{V_{D-2}\over{2\kappa^2_D}}{{2(D-3)}\over\sqrt{\Delta}}m
\sinh\alpha\cosh\alpha,\ \ \ \ \ \ 
\Phi={2\over\sqrt{\Delta}}\tanh\alpha,
\label{thquans}
\end{eqnarray}
where $V_n=2\pi^{{n+1}\over 2}/\Gamma\left({{n+1}\over 2}\right)$ denotes 
the volume of the unit $S^n$.  
These thermodynamic quantities satisfy the first law of black hole 
thermodynamics:
\begin{equation}
dM=TdS+\Phi dQ.
\label{1stthlaw}
\end{equation}

According to the holographic principle, the thermodynamic quantities of a
black hole solution can be identified with the corresponding thermodynamic 
quantities of the holographic dual theory \cite{wit}.  In the case of a 
charged black hole solution, the total energy $E_{tot}=M$ of the dual theory 
can be separated into two parts: the contribution of the supersymmetric 
background (i.e., the zero temperature energy) and that of thermal 
excitation.  Following Refs. \cite{cai,kle2}, we define the zero temperature 
contribution, called the proper internal energy, as
\begin{equation}
E_q={V_{D-2}\over{2\kappa^2_D}}{{4(D-3)}\over\Delta}m\sinh^2\alpha.
\label{prpinte}
\end{equation}
So, the thermal excitation energy is 
\begin{equation}
E=E_{tot}-E_q={V_{D-2}\over{2\kappa^2_D}}(D-2)m.  
\label{thexen}
\end{equation}
If we assume that the holographic dual theory is conformal, then the 
pressure $p=-\left({{\partial E}\over{\partial V}}\right)_{S,Q}$ is given by
\begin{equation}
p={E\over{(D-2)V}}={V_{D-2}\over{2\kappa^2_D}}{m\over V},
\label{prsr}
\end{equation}
where $V$ is the volume of the system.  So, the Casimir energy of the 
holographic dual theory is given by
\begin{equation}
E_c=(D-2)(E_{tot}+pV-TS-\Phi Q)={V_{D-2}\over{2\kappa^2_D}}2(D-2)m.
\label{casenrgmch}
\end{equation}
Since $E_c=2E$, the Cardy-Verlinde formula (\ref{crdvrnlf}) would yield 
$S=0$.  So, we have to consider the modified Cardy-Verlinde formula 
(\ref{zcfwcf}), valid for the case of the ground state with zero conformal 
weight.  It is straightforward to show that thermodynamic quantities $S$, 
$E$ and $E_c$, respectively given in Eqs. 
(\ref{thquans},\ref{thexen},\ref{casenrgmch}), satisfy Eq. (\ref{zcfwcf}),  
provided we identify the length scale of the system as 
\begin{equation}
R=m^{1\over{D-3}}\cosh^{4\over\Delta}\alpha.
\label{lscldchbh}
\end{equation}  
Just like the Schwarzschild black hole case \cite{kle2}, total thermal 
excitation energy (\ref{thexen}) can be put into the following suggestive 
form:
\begin{equation}
E={c\over 6}\pi^2R\left({{2T}\over{D-3}}\right)^2,
\label{2dimrel}
\end{equation}
with the effective central charge $c$ given by 
\begin{equation}
{c\over 6}=2E_cR={{V_{D-2}}\over{2\kappa^2_D}}4(D-2)m^{{D-2}\over{D-3}}
\cosh^{4\over\Delta}\alpha={{D-2}\over\pi}S.
\label{effcchdbh}
\end{equation}
This is the energy-temperature relation of a two-dimensional CFT with the 
characteristic length $R=m^{1\over{D-3}}\cosh^{4\over\Delta}\alpha$ and 
effective temperature $\tilde{T}=2T/(D-3)$ with $T$ given in Eq. 
(\ref{thquans}).  Note, $c\propto S$, which is in accordance with the 
holographic principle.  We can write Eq. (\ref{2dimrel}) also in the form
\begin{equation}
ER={c\over{24}},
\label{2dimrel2}
\end{equation}
which is the ground state energy of a two-dimensional CFT.  The fact that 
thermodynamic quantities of a charged black hole solution (\ref{dilbhsol}) 
satisfy the relation (\ref{2dimrel}) for a two-dimensional CFT supports 
the claim in our previous work \cite{youm1} that thermodynamics of 
higher-dimensional charged black hole solutions can be effectively 
described by two-dimensional theories.  

It is straightforward to show that the modified Cardy-Verlinde formula 
(\ref{zcfwcf}) is satisfied also by multi-charged black hole solutions in 
string theory.  Thermodynamic quantities of such solutions are generally 
given by
\begin{eqnarray}
M&=&{V_{D-2}\over{2\kappa^2_D}}\left[(D-3)m\sum_i\sinh^2\alpha_i
+(D-2)m\right],
\cr
S&=&{{4\pi V_{D-2}}\over{2\kappa^2_D}}m^{{D-2}\over{D-3}}\prod_i
\cosh\alpha_i,\ \ \ \ 
T={{D-3}\over{4\pi}}m^{-{1\over{D-3}}}\prod_i\cosh^{-1}\alpha_i,
\cr
Q_i&=&{V_{D-2}\over{2\kappa^2_D}}m\sinh\alpha_i\cosh\alpha_i,
\ \ \ \ \ \ \ \ \ \ 
\Phi_i=\tanh\alpha_i,
\label{thqmcbh}
\end{eqnarray}
where $i=1,...,4$ for $D=4$ and $i=1,2$ for $D\geq 5$.  These thermodynamic 
quantities satisfy the first law of black hole thermodynamics:
\begin{equation}
dM=TdS+\Phi_idQ_i.
\label{1stthlawmch}
\end{equation}
We define the proper internal energy as
\begin{equation}
E_q={V_{D-2}\over{2\kappa^2_D}}(D-3)m\sum_i\sinh^2\alpha_i.
\label{prpintemch}
\end{equation}
Then, the thermal excitation energy $E=E_{tot}-E_q$ is still given by Eq. 
(\ref{thexen}).  By using the assumed relation $pV=E/(D-2)$, which holds for 
the conformal holographic dual theory, we obtain the following Casimir energy:
\begin{equation}
E_c=(D-2)(E_{tot}+pV-TS-\Phi_iQ_i)={V_{D-2}\over{2\kappa^2_D}}2(D-2)m.
\label{casenrg}
\end{equation}
So, the modified Cardy-Verlinde formula (\ref{zcfwcf}) is satisfied by the 
above thermodynamic quantities, provided we identify the length scale of 
the system as 
\begin{equation}
R=m^{1\over{D-3}}\prod_i\cosh\alpha_i.  
\label{lscmchbh}
\end{equation}
It is straightforward to show that the thermal excitation energy $E$ can also 
be put into the suggestive form (\ref{2dimrel}) with $R$ and $T$ respectively 
given in Eqs. (\ref{lscmchbh},\ref{thqmcbh}) and the effective central charge 
$c$ given by 
\begin{equation}
{c\over 6}=2E_cR={V_{D-2}\over{2\kappa^2_D}}2(D-2)m^{{D-2}\over{D-3}}
\prod_i\cosh\alpha_i={{D-2}\over\pi}S.
\label{effcchmch}
\end{equation}

In our previous work \cite{youm2}, we showed that by compactifying the 
dual-frame action for the Hodge-dual action of Eq. (\ref{action}) on a 
$(D-2)$-sphere with the radius $\hat{\mu}=(m\sinh^2\alpha)^{1/(D-3)}$ 
we obtain the following two-dimensional effective action:
\begin{equation}
S_2={1\over{2\kappa^2_2}}\int d^2x\sqrt{-g}e^{\delta\phi}\left[{\cal R}_g
+\gamma(\partial\phi)^2+\Lambda\right],
\label{2dact}
\end{equation}
where
\begin{equation}
\delta\equiv -{{D-2}\over{D-3}}a,\ \ \ \ 
\gamma\equiv {{D-1}\over{D-2}}\delta^2-{4\over{D-2}},\ \ \ \ 
\Lambda={{D-3}\over{2\hat{\mu}^2}}\left[2(D-2)-{{4(D-3)}\over\Delta}\right].
\label{2dactdefs}
\end{equation}
The Weyl rescaling $\bar{g}_{\mu\nu}=\Phi^{\gamma\over\delta^2}g_{\mu\nu}$ 
($\Phi\equiv e^{\delta\phi}$) brings this action to the form:
\begin{equation}
S_2={1\over{2\kappa^2_2}}\int d^2x\sqrt{-\bar{g}}\left[\Phi{\cal R}_{\bar{g}}
+\Phi^{1-{\gamma\over\delta^2}}\Lambda\right].
\label{2dact2}
\end{equation}
The general static solution to the equations of motion for this 
action in the Schwarzschild gauge is
\begin{eqnarray}
d\bar{s}^2&=&-\left[{\delta^2\over{2\delta^2-\gamma}}\left({x\over\ell}
\right)^{2-{\gamma\over\delta^2}}-2\ell\kappa^2_2\bar{M}\right]d\tau^2+
\left[{\delta^2\over{2\delta^2-\gamma}}\left({x\over\ell}\right)^{2-
{\gamma\over\delta^2}}-2\ell\kappa^2_2\bar{M}\right]^{-1}dx^2,
\cr
\Phi&=&{x\over\ell},
\label{gen2dsol}
\end{eqnarray}
where $\ell\equiv1/\sqrt{\Lambda}$ and $\bar{M}$ is the ADM mass of the 
solution.  Thermodynamic quantities of the solutions are
\begin{equation}
\bar{M}={\delta^2\over{2\delta^2-\gamma}}{1\over{2\ell\kappa^2_2}}
\Phi^{2-{\gamma\over\delta^2}}_H,\ \ \ \ \ \ \ \ 
\bar{T}={1\over{4\pi\ell}}\Phi^{1-{\gamma\over\delta^2}}_H,\ \ \ \ \ \ \ \ 
\bar{S}={{2\pi}\over\kappa^2_2}\Phi_H,
\label{2dbhthq}
\end{equation}
where $\Phi_H$ is the value of $\Phi$ at the event horizon, i.e., 
$\bar{g}_{\tau\tau}(\Phi_H)=0$.  If we assume the two dimensional system 
under consideration to be conformal, then $\bar{p}\bar{V}=\bar{M}$, so the 
Casimir energy is given by
\begin{equation}
\bar{M}_c=\bar{M}+\bar{p}\bar{V}-\bar{T}\bar{S}={\gamma\over{2\delta^2-
\gamma}}{1\over{2\ell\kappa^2_2}}\Phi^{2-{\gamma\over{\delta^2}}}_H.
\label{2dcasen}
\end{equation}
These thermodynamic quantities satisfy the following Cardy formula \cite{car} 
for a two-dimensional CFT:
\begin{equation}
\bar{S}=2\pi\sqrt{{c\over 6}\left(L_0-{c\over{24}}\right)},
\label{2dcardf}
\end{equation}
with $L_0=2\ell\bar{M}$ and $c/6=4\ell\bar{M}_c$, provided $\gamma=\delta^2$.  
This is in agreement with the result in our previous work \cite{youm2} 
that the action (\ref{2dact2}) is conformal if $\gamma=\delta^2$.

\end{document}